\documentclass[aps,prb,10pt,twocolumn,showpacs,amsmath,amssymb,floatfix,superscriptaddress]{revtex4-2}

\usepackage{graphicx}
\usepackage{color}
\usepackage{bm}
\usepackage{braket}
\usepackage[export]{adjustbox}
\usepackage[colorlinks=true,allcolors=blue]{hyperref}

\newcommand{\ft}[1]{{\cal #1}}
\newcommand{\lb}{\left[}
\newcommand{\rb}{\right]}
\newcommand{\lp}{\left(}
\newcommand{\rp}{\right)}

\newcommand{\eps}{\varepsilon}
\renewcommand{\epsilon}{\varepsilon}
\renewcommand{\tilde}{\widetilde}
\renewcommand{\phi}{\varphi}
\newcommand{\zero}{{(0)}}

\renewcommand{\vec}[1]{{\bf #1}}
\newcommand{\vk}{\vec{k}}
\newcommand{\vq}{\vec{q}}
\renewcommand{\vr}{\vec{r}}

\def\nn{\nonumber\\}

\def\afflux{Department of Physics and Materials Science, University of Luxembourg,\\ L-1511 Luxembourg, Luxembourg}
\begin{document}

\title{Hydrodynamic Navier-Stokes equations in two-dimensional\\ systems with Rashba spin-orbit coupling}

\author{Edvin G. Idrisov}
\affiliation{\afflux}
\affiliation{Abrikosov Center for Theoretical Physics, MIPT, 141701 Institutskii per., Dolgoprudnyi, Russia}

\author{Eddwi H. Hasdeo}
\affiliation{\afflux}
\affiliation{Research Center for Quantum Physics, National Research and Innovation Agency (BRIN),\\ South Tangerang 15314, Indonesia}

\author{Byjesh N. Radhakrishnan}
\affiliation{\afflux}

\author{Thomas L. Schmidt}
\affiliation{\afflux}
\affiliation{School of Chemical and Physical Sciences, Victoria University of Wellington,\\ P.O. Box 600, Wellington 6140, New Zealand}

\date{\today}
	
\begin{abstract}
We study a two-dimensional (2D) electron system with a linear spectrum in the presence of Rashba spin-orbit (RSO) coupling in the hydrodynamic regime. We derive a semiclassical Boltzmann equation with a collision integral due to Coulomb interactions in the basis of the eigenstates of the system with RSO coupling. Using the local equilibrium distribution functions, we obtain a generalized hydrodynamic Navier-Stokes equation for electronic systems with RSO coupling. In particular, we discuss the influence of the spin-orbit coupling on the viscosity and the enthalpy of the system and present some of its observable effects in hydrodynamic transport.
\end{abstract}

\maketitle

\section{Introduction}
\label{Sec:I}
Hydrodynamic behavior of electrons in metals was first predicted in 1963 by Gurzhi~\cite{gurzhi1963}. It became clear that the hydrodynamic regime in conductors can be reached when the electron-electron scattering time $\tau_{ee}$ is the shortest time scale compared with the electron-impurity ($\tau_{ei}$) and electron-phonon ($\tau_{eph}$) scattering times. At that time it was a challenge to fabricate samples clean enough to satisfy this condition, so the first experimental observation of the hydrodynamic regime was demonstrated only in 1995 by de Jong and Molenkamp~\cite{DeJong1995}.
It is well known that the scattering times $\tau_{ee}$, $\tau_{ei}$ and $\tau_{eph}$ strongly depend on temperature~\cite{Abrikosov}. Electron-impurity scattering processes are most essential at low temperatures, whereas the electron-phonon mechanism becomes dominant for high temperatures. In certain materials, a hydrodynamic regime is thus reached at intermediate temperatures if they are sufficiently clean.

In the recent past, the technological progress in the fabrication of 2D materials has reignited the interest in electron hydrodynamics~\cite{PoliniGeim}. In particular, monolayer graphene with its linear Dirac-like spectrum has become a fruitful experimental platform to investigate hydrodynamic transport~\cite{Zaanen1026}. It was shown that the hydrodynamic regime in clean graphene can be realized at temperatures on the order of $100$K~\cite{Derek}. Many peculiar transport properties have been demonstrated in the hydrodynamic regime~\cite{Lucas2018}. For instance, Johnson thermometry measurements show a significant increase in the thermal conductivity and the breakdown of the Wiedemann-Franz law in graphene~\cite{Crossno2016}. This is possible due to a decoupling of charge and heat currents within the hydrodynamic regime and can be regarded as a signature of a Dirac fluid~\cite{Crossno2016}. Viscous electron flow through constrictions in graphene has revealed superballistic behavior~\cite{Guo3068}, i.e., a conductance exceeding the maximum conductance possible for ballistic electrons in the same geometry~\cite{KrishnaKumar2017}. Another signature of collective viscous behavior is the non-local negative resistance and the enhancement of the thermoelectric power in a graphene strip~\cite{Levitov2016,Ghahari}. A viscous flow of the Dirac fluid in graphene was also confirmed using a quantum spin magnetometer~\cite{Ku2020}. Besides graphene, the electron hydrodynamics regime has been theoretically investigated and in certain cases also experimentally confirmed in 2D anomalous Hall materials~\cite{Hasdeo2021}, in anisotropic materials~\cite{Varnavides2020}, in Coulomb drag geometries~\cite{Apostolov2014,Chen2015,Hasdeo2023,Kavokine2023}, in Weyl semimetals~\cite{Zhu2022}, in gallium arsenide~\cite{keser2021}, and in 2D electron gases with spin-orbit interaction~\cite{Doornenbal2019,matsuo2020spin}.

Nowadays, it has become possible to fabricate a plethora of hybrid systems based on graphene~\cite{Chen}, for instance by combining graphene with adatoms~\cite{Castronet2009}, two-dimensional transition metal dichalcogenides~\cite{Gmitra2015}, or thin metallic substrates~\cite{Dedkov2008,Marchenko2012,Marchenko2013}, which make it possible to manipulate the spin degree of freedom and are promising for spintronics~\cite{Han2014}. Importantly, even in these hybrid structures, graphene with globally induced spin-orbit coupling remains essentially free from defects and impurities. The intrinsic spin-orbit coupling in graphene is weak, but in these hybrid structures, proximity-induced spin-orbit interaction can be large and can change the electronic band structure substantially. Intrinsic spin-orbit coupling opens a gap at the $K$ point and is related with the pseudospin inversion asymmetry, whereas Rashba spin-orbit (RSO) coupling preserves the gapless nature of graphene~\cite{KaneOne2005,KaneTwo2005}. Typically the RSO interaction appears due to the structural inversion asymmetry brought about by the substrate or adatoms~\cite{Han2014}. It is worth mentioning that the induced RSO coupling can reach larger values in other two-dimensional materials such as silicene, germanene, stanene, phosphorene, arsenene, antimonene, and bismuthene~\cite{Kurpas2019}.

In this work we study the effect of RSO coupling on hydrodynamic transport in graphene. We allow the spin-orbit coupling to be on the same order as the hydrodynamic temperature and assume that the electron-electron scattering time remains the shortest time scale. The intrinsic spin-orbit coupling can be tuned to small values~\cite{Han2014,Gmitra2015}, so we ignore its influence on transport properties and assume that the system remains gapless. In order to derive the hydrodynamic equations, we use the kinetic (Boltzmann) equation in the diagonal basis of the unperturbed Hamiltonian, which includes the Dirac spectrum and the RSO interaction. The collision integral on the right-hand side of the kinetic equation accounts for two-particle scattering. The RSO coupling results in the appearance of so-called Dirac factors in the two-body electron-electron interaction Hamiltonian. For our derivation we mainly follow Ref.~[\onlinecite{Narozhny2019}], where the necessary calculations were performed for pristine graphene.

The rest of this article is organized as follows. In Sec.~\ref{Sec:II}, we introduce the Hamiltonian of the 2D system under consideration. In Sec.~\ref{Sec:III}, we provide the Boltzmann equation for the system taking into account the spin-orbit split conduction bands. In Sec.~\ref{Sec:IV}, using the kinetic equation and thermodynamic relations, we derive the generalized hydrodynamic Navier-Stokes equation. Finally, we present our conclusions in Sec.~\ref{Sec:V}. The details of the calculations and additional information are presented in the Appendices. Throughout the paper, we set $e=\hbar=k_B=1$.

\section{System Hamiltonian}
\label{Sec:II}

In this section, we derive the Hamiltonian of the system under consideration. Specifically, we consider a single layer of graphene on a substrate which enhances the RSO coupling~\cite{Gmitra2015}. The first-quantized one-body Hamiltonian of graphene near the $K$ point in the momentum representation is given by
\begin{equation}
\label{Pristine graphene's Hamiltonian}	
H_0=v(\sigma_x k_x+\sigma_y k_y),
\end{equation}
where $v \simeq 10^{6}$m/s is the Fermi velocity for spinless electrons, and $\sigma_{x,y}$ are the pseudospin Pauli matrices acting on the two-dimensional space of sublattices. This Hamiltonian describes a linear gapless spectrum near a given Dirac point. The large separation in momentum space suppresses (inter-valley) scattering between the $K$ and $K'$ points, so we focus on a single Dirac cone in the following.

A broken structural inversion symmetry results in the RSO interaction Hamiltonian
\begin{equation}
\label{Rashba spin-orbit interaction Hamiltonian}
H_{R}=\lambda_R(\sigma_x s_y-\sigma_y s_x),
\end{equation}
where $\lambda_R$ denotes the strength of the RSO coupling and $s_{x,y}$ are the Pauli matrices corresponding to the electron spin. It is worth mentioning that the RSO term in graphene does not depend explicitly on momentum in contrast to the general case for two-dimensional electron gases~\cite{Shytov2006,Doornenbal2019}. In order to simplify the calculation, we neglect two additional perturbations which can be present in graphene, namely a staggered sublattice potential and the intrinsic spin-orbit coupling. Passing on to a second-quantized description, the graphene Hamiltonian in the presence of the RSO term has the form
\begin{equation}
\label{Zero Hamiltonian in second quantization}
H=H_0+H_R=\sum_{n\vk} \epsilon_{n\vk} c^{\dagger}_{n\vk}c_{n\vk}.	
\end{equation}
The eigenstates of the single-particle Hamiltonian~(\ref{Zero Hamiltonian in second quantization}) are denoted by $|n\vk\rangle$, where $n$ denotes the band index and $\vk=(k_x,k_y)$ is the wave vector in 2D momentum space. The annihilation and creation operators of an electron with momentum $\vk$ in band $n$ satisfy the standard anti-commutation relations, $\{c^{\dagger}_{n\vk},c_{n^{\prime} \vk^{\prime}}\}=\delta_{nn^{\prime}}\delta_{\vk \vk^{\prime}}$. The RSO coupling leaves the spectrum gapless at $|\vk|=0$ and one finds the dispersion relations for the four bands, namely
\begin{equation}
\label{Spectrum of zero Hamiltonian with spin-orbit coupling}
\epsilon_{n\vk}=
\begin{cases}
	     \mp\lambda_R-\sqrt{v^2 \vk^2+\lambda^2_R} & \quad n=v_1,v_2, \\
	      	\mp\lambda_R+\sqrt{v^2 \vk^2+\lambda^2_R} & \quad n=c_1,c_2.
\end{cases}	
\end{equation}
where the band indices $c_{1,2}$ and $v_{1,2}$ refer to the spin-split conductance and valence bands, respectively.

The dominant scattering mechanism to reach the hydrodynamic regime is the electron-electron interaction~\cite{Lucas2018,Narozhny2019}. In second quantization, the electron-electron interaction in the eigenbasis of the diagonal Hamiltonian~(\ref{Zero Hamiltonian in second quantization}) can be written as
\begin{align}
\label{Electron-electron interaction Hamiltonian in second quantization}
H_{ee}&=\frac{1}{2S}\sum_{\substack{n_1 n_3 \\
\vk_1 \vk_3}} \sum_{\substack{n_2 n_4 \\
\vk_2 \vk_4}}\sum_{\vq}V_{\vq} F^{n_1 n_3}_{\vk_1 \vk_3} F^{n_2 n_4}_{\vk_2 \vk_4} \notag \\
&\times \delta_{\vk_1+\vq, \vk_3} \delta_{\vk_2-\vq, \vk_4} c^{\dagger}_{n_3 \vk_3}	c^{\dagger}_{n_4 \vk_4}c_{n_2 \vk_2} c_{n_1 \vk_1},	
\end{align}
where $S$ is the 2D system volume. The interaction potential $V_{\vq}$ is given by the Fourier transform of the real-space interaction potential, $V(\vr_1-\vr_2)=(1/S)\sum_{\vq}V_{\vq}e^{i\vq(\vr_1-\vr_2)}$, and has the symmetry $V_{-\vq}=V_{\vq}$. The Dirac factors in the presence of RSO interaction are represented by the following matrix element (see App.~\ref{Sec:A} for details)
\begin{align}
\label{Dirac factor in presence of spin-orbit coupling}
    F^{nn^{\prime}}_{\vk \vk^{\prime}}
&=
    \frac{1}{2} \xi^n_{\vk} \xi^{n^{\prime}}_{\vk^{\prime}}
    \Big[
    1 + nn^{\prime} g^{n}_{\vk,1}(g^{n^{\prime}}_{\vk^{\prime},1})^{\ast} \notag \\
&+
    g^{n}_{\vk,2}(g^{n^{\prime}}_{\vk^{\prime},2})^{\ast} +
    nn^{\prime}g^{n}_{\vk,3}(g^{n^{\prime}}_{\vk^{\prime},3})^{\ast}\Big],	
\end{align}
where
\begin{align}
    \xi^{n}_{\vk} &= \left[1+(\zeta^n_{\vk})^2\right]^{-1/2}, &
    \zeta^n_{\vk} &= \epsilon_{n\vk}/(v|\vk|), \notag \\
    g^n_{\vk,1} &= i\zeta^n_{\vk}e^{-i\theta_{\vk}}, &
    g^n_{\vk,2} &= \zeta^n_{\vk}e^{-i\theta_{\vk}}, \notag \\
    g^n_{\vk,3} &= ie^{-2i\theta_{\vk}},
\end{align}
and $\theta_{\vk}$ denotes the angle of the 2D vector $\vk$ with respect to the $k_x$ axis. The superscript $n$ denotes the band and $n=+1$ ($n=-1$) corresponds to the $v_1, c_1$ ($v_2,c_2$) bands. It is worth mentioning that $(F^{nn^{\prime}}_{\vk \mathbf{k^{\prime}}})^{\ast}=F^{n^{\prime} n}_{\mathbf{k^{\prime}} \vk}$, which guarantees the hermiticity of the electron-electron interaction Hamiltonian.

\section{Boltzmann equation}
\label{Sec:III}
The kinetic equation can be obtained by applying the Keldysh technique to the total Hamiltonian given in previous section and applying perturbation theory and a semiclassical approximation. The details of these steps are standard and explained in many reviews~\cite{Rammer1986,Kita2010,Arseev2015}. Choosing a chemical potential above the charge neutrality point, we need to consider only the two conduction bands. We denote the corresponding semiclassical distribution functions as $f_{n_1 \vk_1} \equiv f_{n_1 \vk_1}(\vr,t)$, where $n_1=+1$ ($n_1 = -1$) corresponds to the lower (upper) conduction band $c_1$ ($c_2$). The exact form of the distribution functions obtained from the Keldysh Green's functions is provided in Refs.~[\onlinecite{Rammer1986}] and [\onlinecite{Kita2010}]. In our effective two-band system, the Boltzmann equation for the distribution functions can be written as
\begin{equation}
\label{Kinetic equation}
\frac{\partial f_{n_1\vk_1}}{\partial t}+\frac{\partial \epsilon_{n_1\vk_1}}{\partial \vk_1}\frac{\partial f_{n_1 \vk_1}}{\partial \vr}-\frac{\partial \varphi}{\partial \vr} \frac{\partial f_{n_1 \vk_1}}{\partial \vk_1}=\mathcal{I}[f_{n_1\vk_1}],
\end{equation}
where $\varphi(\vr,t)$ is an external field applied to the system and $\mathcal{I}[f_{n_1 \vk_1}]$ is the electron-electron collision integral, which is a functional of the distribution functions. Our main interest is in the hydrodynamic regime where electron-electron interactions dominate. For simplicity, we therefore omit the scattering integrals related to electron-hole recombination as well as the scattering integrals related to impurity and phonon scattering. In the two-particle scattering approximation the scattering integral has the form
\begin{equation}
\label{The initial form of scattering integral}
\mathcal{I}[f_1]=-2\pi\mathcal{N}\sum_{23 4}W^{3 4}_{12}\left[\mathcal{G}^{3 4}_{12}-\mathcal{G}_{3 4}^{12}\right],
\end{equation}
where we used the shorthand notation $1\equiv(n_1\vk_1)$ etc. The outgoing and incoming fluxes are denoted by $\mathcal{G}^{3 4}_{12}=f_1f_2(1-f_{3})(1-f_4)$ and the valley degeneracy factor $\mathcal{N}=2$. The two-particle scattering rate is given by Fermi's golden rule
\begin{equation}
\label{Fermi's golden rule}
W^{3 4}_{12}=|\langle 3 4|H_{ee}|12\rangle_c|^2\delta(E_i-E_f),
\end{equation}
where $i=|12\rangle$ and $f=|34\rangle$ denote, respectively, the initial and final states of the scattering process. Moreover, $E_i=\epsilon_{n_1\vk_1}+\epsilon_{n_2\vk_2}$ is the initial state energy, the subscript $c$ means that only connected diagrams are taking into account, and the delta function is a consequence of energy conservation. Applying Wick's theorem, we obtain the following expression for the two-particle transition matrix element
\begin{align}
\label{Transition matrix element}
    \langle 3 4|H_{ee}|12\rangle_c
&=
    \frac{1}{S}\Big[V_{\vk_1-\vk_4}F^{n_1n_4}_{\vk_1 \vk_4}F^{n_2n_3}_{\vk_2\vk_3}  \\
&-
    V_{\vk_1-\vk_3}F^{n_1n_3}_{\vk_1 \vk_3}F^{n_2n_4}_{\vk_2\vk_4}\Big]\delta_{\vk_1+\vk_2,\vk_3+\vk_4},	\notag
\end{align}
where the Kronecker-delta ensures momentum conservation in the two-particle collision process. It is worth mentioning that in Ref.~[\onlinecite{Narozhny2019}] only the first term of Eq.~(\ref{Transition matrix element}) is kept, thus ignoring the interference effects. In this case, at zero spin-orbit coupling, our results reproduce those of Ref.~[\onlinecite{Narozhny2019}], namely $|\langle 3 4|H_{ee}|12\rangle_c|^2=|V_{\vk_1-\vk_3}|^2 \Theta^{n_1n_3}_{\vk_1\vk_3} \Theta^{n_2n_4}_{\vk_2\vk_4}\delta_{\vk_1+\vk_2,\vk_3+\vk_4}$ with Dirac factors for pristine graphene $\Theta^{nn^{\prime}}_{\vk\vk^{\prime}}=(1+nn^{\prime} \mathbf{e}_{\vk} \cdot \mathbf{e}_{\vk^{\prime}})/2$, where $\mathbf{e}_{\vk}=\vk/|\vk|$ is a unit vector in the direction of the momentum.

The strong interactions cause the electrons to relax on a short time scale to local equilibrium distributions
\begin{equation}
\label{Local equilibrium distribution function}
f^{\text{eq}}_{m\vk}(\vr,t)=\left\{1+\exp\left[\frac{\epsilon_{m\vk}-\mu(\vr,t)-\mathbf{u}(\vr,t)\cdot \vk}{T(\vr,t)}\right]\right\}^{-1},
\end{equation}
where $\mu$ is the chemical potential and $\mathbf{u}$ and $T$ are the local drift velocity and the temperature, respectively. This form of the distribution function is a general consequence of Boltzmann's H-theorem, which states that in local equilibrium the entropy production must vanish. The latter is defined as
 \begin{widetext}
\begin{equation}
\left[\frac{\partial S}{\partial t}\right]_{\text{coll.}}=\frac{1}{4}\sum_{\{n\}} \int \frac{d\vk_1}{2\pi}\int \frac{d\vk_2}{2\pi} \int \frac{d\vk_3}{2\pi} \int \frac{d\vk_4}{2\pi}|\langle 3 4|H_{ee}|1 2\rangle_c|^2\delta(\mathbf{k}_i-\mathbf{k}_f)\delta(E_i-E_f)\left[\mathcal{G}^{12}_{3 4}-\mathcal{G}_{12}^{3 4}\right]\log\left[\frac{\mathcal{G}^{12}_{3 4}}{\mathcal{G}_{12}^{3 4}}\right],
\end{equation}
\end{widetext}
where $\{n\}=(n_1,n_2,n_3,n_4)$ and $\mathbf{k}_{i,f}$, and $E_{i,f}$ denote the total momentum and energy, respectively, of the initial and final states. Analogous expressions have been already derived for different systems, in particular for Fermi liquids and (non-linear) Luttinger liquids~\cite{Kita2010,Idrisov2019}. Using the property $(\mathcal{G}^{12}_{3 4}-\mathcal{G}_{12}^{3 4})\log (\mathcal{G}^{12}_{3 4}/\mathcal{G}_{12}^{3 4})>0$, one obtains indeed $[\partial S/\partial t]_{\text{coll.}} \geq 0$ and the local distribution function~(\ref{Local equilibrium distribution function}) can be deduced from the zero entropy production condition, i.e., $[\partial S/\partial t]_{\text{coll.}}=0$. The kinetic equation~(\ref{Kinetic equation}) and the local distribution function~(\ref{Local equilibrium distribution function}) will be used in the next section to derive the hydrodynamic equations.

\section{Navier-Stokes equation}
\label{Sec:IV}
In principle, hydrodynamic equations can be formulated on the basis of the conservation laws of particle number, momentum and energy~\cite{Landau}. Moreover, it is known that the hydrodynamic equations can be derived explicitly from the kinetic equation, as was done in Ref.~[\onlinecite{Narozhny2019}] for graphene without spin-orbit coupling. We proceed along similar lines and start by considering the continuity equations and conservation laws arising from the Boltzmann equation~(\ref{Kinetic equation}).

The continuity equation for the particle number can be obtained by integrating Eq.~(\ref{Kinetic equation}) over momentum. In this case the right-hand side of the kinetic equation vanishes and introducing the particle number and current
\begin{equation}
\label{Particle number and current}
\begin{aligned}
n(\vr,t) &= \sum_{m} \int \frac{d\vk}{(2\pi)^2}f_{m\vk}(\vr,t), \\
\mathbf{j}(\vr,t) &=\sum_m \int \frac{d\vk}{(2\pi)^2} \mathbf{v}_{m\vk} f_{m\vk}(\vr,t),
\end{aligned}	
\end{equation}
where $\mathbf{v}_{m\vk}=\partial \epsilon_{m\vk}/\partial \vk=v^2 \vk/\sqrt{v^2 \vk^2+\lambda^2_R}$,
one arrives at the continuity equation for the particle density
\begin{equation}	
\label{Continuity equation for particle number}
\frac{\partial n}{\partial t}+\nabla \cdot \mathbf{j}=0.
\end{equation}
Similarly, introducing the imbalance particle number and the imbalance current
\begin{equation}
	\label{Particle number and current_I}
	\begin{aligned}
	n_I(\vr,t) &= \sum_{m} \int \frac{d\vk}{(2\pi)^2}m f_{m\vk}(\vr,t), \\
    \mathbf{j}_I(\vr,t) &=\sum_m \int \frac{d\vk}{(2\pi)^2} m\mathbf{v}_{m\vk} f_{m\vk}(\vr,t),
	\end{aligned}	
\end{equation}
one straightforwardly obtains the continuity equation for imbalance quantities
\begin{equation}
\label{Continuty equation for imbalance particle and current}
\frac{\partial n_I}{\partial t}+\nabla \cdot \mathbf{j}_I=0.
\end{equation}
To obtain the continuity equation for the energy density $n_E(\vr,t)$ and the energy (heat) current $\mathbf{j}_E(\vr,t)$, both sides of Eq.~(\ref{Kinetic equation}) are multiplied by $\epsilon_{n\vk}$ and integrated over momentum. As a result, one finds
\begin{equation}
\label{Continuity equation for energy density}
\frac{\partial n_E}{\partial t}+\nabla \cdot \mathbf{j}_E=\mathbf{E} \cdot \mathbf{j},
\end{equation}
where the Joule heat term on the right-hand side contains the electric field $\mathbf{E}=-\nabla \varphi$.

In order to derive the continuity equation for the momentum density, one needs to multiply the kinetic equation by the components of $\vk$ and integrate over momentum. Therefore, the continuity equation in this case has the form
\begin{equation}
\label{Continuity equation for momentum density}
\frac{\partial n^{i}_{\vk}}{\partial t}+\sum_{j} \frac{\partial \Pi^{ij}}{\partial j} 	=nE_{i}, \quad i, j \in \{ x,y\},
\end{equation}
where the momentum density and momentum flux tensor are given by
\begin{equation}
\label{Momentum density and momentum flux tensor}
\begin{aligned}
n^{i}_{\vk}(\vr,t)&=\sum_m \int \frac{d\vk}{(2\pi)^2} k_{i}f_{m\vk}(\vr,t),\\
\Pi^{ij}(\vr,t)&=\sum_m \int \frac{d\vk}{(2\pi)^2} k_{i} v^j_{m\vk} f_{m\vk}(\vr,t).
\end{aligned}
\end{equation}

Next we derive the macroscopic expressions which relate the currents with the densities and the drift velocity $\mathbf{u}(\vr,t)$. Using the local distribution functions~(\ref{Local equilibrium distribution function}) and the definitions of the densities and the currents, one can show the following relations (see App.~\ref{Sec:B} for details)
\begin{align}
\label{Macroscopic equations for densities and currents}
    \mathbf{j}   &= n\mathbf{u}, \quad
    \mathbf{j}_I = n_I\mathbf{u}, \quad
    \mathbf{j}_E = W\mathbf{u}, \notag \\
    \mathbf{n}_{\vk} &= v^{-2}(W+\lambda_R n_I)\mathbf{u}, \\
    \Pi^{i j} &= P \delta_{i j}+v^{-2}(W+\lambda_R n_I)u_{i}u_{j}+\Pi^{ij}_d,
\end{align}
where $W$ is the enthalpy per volume (it has a dimension of pressure and we will call it enthalpy instead of enthalpy density below) and $P$ is the pressure. They are related with each other through the energy density~\cite{Landau}, namely
\begin{equation}
\label{Enthalpy}
W=n_E+P,
\end{equation}
and, according to thermodynamics, the pressure in the local equilibrium state is given by
\begin{equation}
\label{Pressure}
P=\frac{1}{\beta}\sum_n \int \frac{d\vk}{(2\pi)^2} \log[1+e^{-\beta(\epsilon_{n\vk}-\mu-\mathbf{u} \cdot \vk)}],	
\end{equation}
where $\beta=1/T$ is a local inverse temperature.
Finally, $\Pi^{i j}_d$ in Eq.~\eqref{Macroscopic equations for densities and currents} is a dissipative contribution which is related to the electron shear viscosity (see App.~\ref{sec:collision}),
\begin{equation}
\Pi_d=-\frac{ \nu_0 \tilde W}{v^2} \lb  (\partial_x u_x-\partial_y u_y)\tau_z- (\partial_x u_y +\partial_y u_x)\tau_x\rb,\label{eq:stress2}
\end{equation}
where $\tau_{x,z}$ are Pauli matrices, $\tilde{W}=W+\lambda_R n_I$, and $\nu_0$ is the (static) kinematic viscosity
\begin{equation}
\nu_0=\frac{v_F^2}{4 g_F\gamma_{ee}},
\end{equation}
with $\gamma_{ee}$ being the electron-electron scattering rate, $v_F=(2\pi)^{-1}\sum_{m} \int k\, dk\, (\partial_k \eps_{mk}) \delta(\mu -\eps_{mk} )$ is the band-averaged Fermi velocity and $g_F=(2\pi)^{-1}\sum_{m} \int k\, dk \delta(\mu -\eps_{mk} )$ is the density of states at Fermi surface.

First, we investigate how spin-orbit coupling affects $\gamma_{ee}$. In order to evaluate the scattering integral, we linearize the Boltzmann equation~\eqref{Kinetic equation} by assuming small $\vec u$ and a nonequilibrium distribution function which is localized near the Fermi level. We refer to App.~\ref{sec:collision} for the detailed derivation.

The electron-electron (e-e) interactions conserve particle number, energy and momentum which, respectively, correspond to the zeroth and the first-harmonic angular function of the nonequilibrium distribution [see Eq.~\eqref{eq:angular}]. Thus, the relaxation originates from second harmonics or higher. It is known that even harmonics decay faster than the odd ones~\cite{ledwidth19} and here we assume that the viscosity only comes from the second harmonic. As we discuss in App.~\ref{sec:collision}, we neglect interband scattering. We then obtain $\gamma_{ee}=\sum_{m}\gamma_2^m$ where $\gamma_2^m$ is the e-e scattering rate of the second harmonic nonequilibrium distribution of the conduction band $m$. Note that the electron wave functions inside the Dirac factor of the Coulomb matrix elements~\eqref{Dirac factor in presence of spin-orbit coupling} play a crucial role in determining the dependence of $\gamma_2^m$ on $\lambda_R$. Eventually one finds that $\gamma_2^m$ scales with $T^2$ as
\begin{align}
\gamma_2^m=\frac{32}{3} \frac{e^4  T^2}{ v^4} \tilde \gamma^m(\tilde\lambda_R, \tilde d), \label{eq:gamma_2}
\end{align}
where
\begin{align}
&
    \tilde \gamma^\pm(\tilde\lambda_R, \tilde d)
=
    \int_0 ^{2} d\tilde q\ \frac{\tilde q-\tilde q^3/4}{(\tilde q+\tilde d)^2} \Bigg\{\sqrt{4-\tilde q^2} \label{eq:gtilde}\\
&-
    \tilde q \displaystyle \left[ \tan^{-1} \lp \frac{\sqrt{4-\tilde q^2}}{\tilde q}\rp\mp \tan^{-1} \lp \frac{\tilde \lambda_R}{\tilde q}\sqrt{\frac{4-\tilde q^2}{1+ \tilde \lambda_R^2}}\rp \right] \Bigg\}^2 \notag ,
\end{align}
is a dimensionless function whose value depends on $\tilde\lambda_R=\lambda_R/(vk_F)$ and the (dimensionless) inverse screening length $\tilde d = d/k_F$, where $k_F$ is the Fermi wavevector. For this derivation, we have used a screened Coulomb potential with $V_q=2\pi e^2/(q+d)$. We note that the $\pm$ sign in Eq.~\eqref{eq:gtilde} originates from the electron wavefunction which encodes the different dispersions of two conduction bands.  Using as parameters $\lambda_R=0.01$~eV, $\mu=0.1$~eV, $d/k_F=1$, $v=10^6$~m/s, and $T=100$~K, we obtain a typical  static viscosity $\nu_0\approx 2\ \mathrm{nm^2/fs}$.

\begin{figure}[t]
    \includegraphics[width=6cm,left]{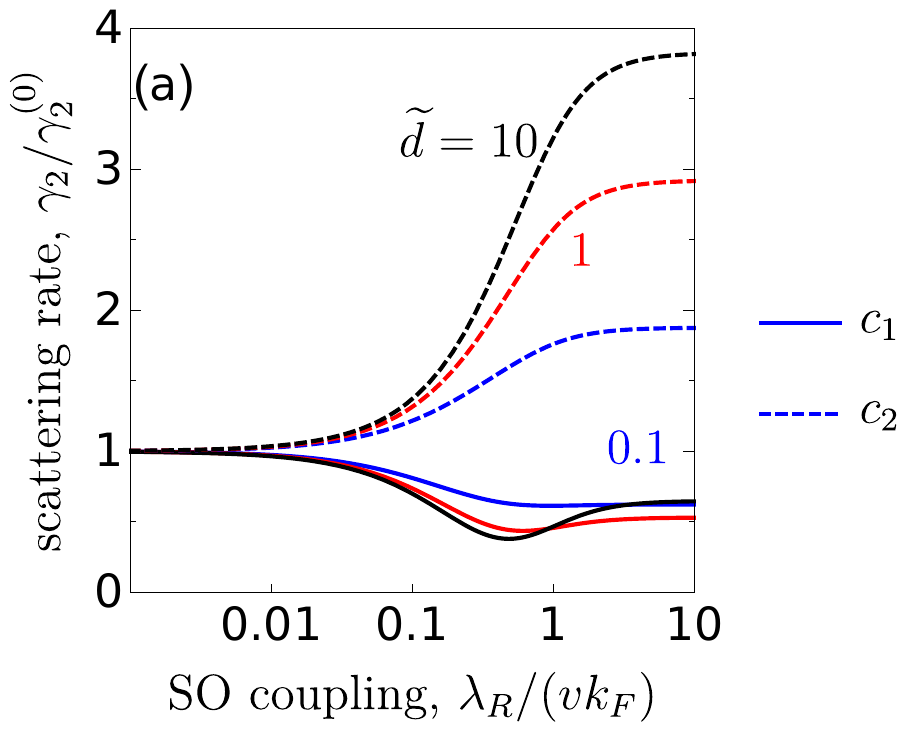}
     \includegraphics[width=6.5cm,left]{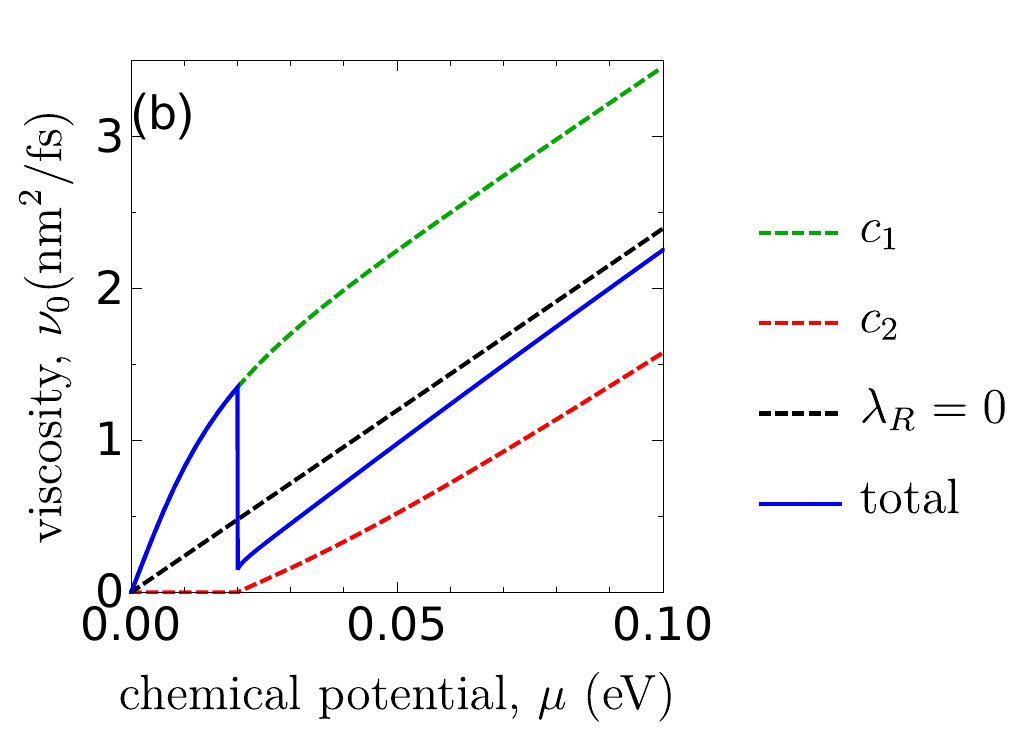}
    \caption{(a) Electron-electron scattering rate of the $c_1$ band (dash-dotted lines), the $c_2$ band (dashed lines), and their average (solid lines).  Here, $\gamma_2^\zero$  denotes the  values of the scattering rate for $\lambda_R=0$. (b) Static viscosity as a function of chemical potential for $\tilde d=d/k_F=1$ and $\lambda_R=0.01\ {\rm eV}$ .  }
    \label{fig:gamma}
\end{figure}

We illustrate the dependence of $\gamma_2^m$ on $\lambda_R$ and $d$ in Fig.~\ref{fig:gamma}(a). Solid lines denote the contribution from the $c_1$ ($m=-1$) band, the dashed lines are from the $c_2$ ($m=+1$) band. As $\lambda_R$ increases, the scattering rate can be non-monotonic and this non-trivial behavior originates from the electron wave functions. We note that the scaled values $\tilde \lambda_R$ and $\tilde d$ can in principle be different in the two bands as they will have different Fermi momenta $k_F$. We also plot the static viscosity as a function of $\mu$ in Fig.~\ref{fig:gamma}(b) for $\lambda_R=0.01\ {\rm eV}$. We also compare the contribution of $c_1$ (green line) and $c_2$ (red line) to the viscosity as well as the viscosity of graphene without RSOC (black line).

At $\mu<2\lambda_R$, only $c_1$ contributes and the viscosity value is larger than graphene without RSOC because $\gamma_2^-/\gamma_2^{(0)}$ is less than one for small $\mu$ (or large $\lambda_R/(vk_F)$) [see Fig.~\ref{fig:gamma}(a)]. At $\mu>2\lambda_R$, the viscosity drops due to the additional contribution of scattering rate from $c_2$ whose value of   $\gamma_2^+/\gamma_2^{(0)}$ is larger than unity. At very large $\mu$, the viscosity gets closer to the value of graphene without RSOC which corresponds to small $\lambda_R/(vk_F)$ in Fig.~\ref{fig:gamma}(a). 

Now, we have all ingredients to derive the Navier-Stokes equation. We find
\begin{align}
\label{Derivation of Euler equation steps1}
\partial_t \mathbf{n}_{\vk}&=v^{-2} \left(\tilde{W} \partial_t \mathbf{u}+\mathbf{u} \partial_t \tilde{W} \right), \\
\frac{\partial  \Pi^{i j}}{\partial j} &= \frac{\partial P}{\partial i} +\frac{1}{v^2}\left\{\tilde{W}\lb \lp \mathbf{u} \cdot \nabla\rp u_{i} - \nu \nabla^2 u_i \rb +u_{i}\nabla \cdot (\tilde{W} \mathbf{u})\right\}. \notag
\end{align}
Then, from Eqs.~(\ref{Enthalpy}) and (\ref{Continuty equation for imbalance particle and current}) one obtains
\begin{equation}
\label{Derivation of Euler equation steps2}
\partial_t(\tilde{W}-P)+\nabla \cdot (\tilde{W} \mathbf{u})=\mathbf{E} \cdot \mathbf{j}.
\end{equation}
Finally, substituting Eqs.~(\ref{Derivation of Euler equation steps1}) and (\ref{Derivation of Euler equation steps2}) into Eq.~(\ref{Continuity equation for momentum density}), we arrive at the hydrodynamic Navier-Stokes equation
\begin{align}
\label{Euler equation}
&W(\partial_t+\mathbf{u} \cdot \nabla -\nu \nabla^2) \mathbf{u}+v^2 \nabla P+\mathbf{u}\partial_t P+(\mathbf{E} \cdot \mathbf{j})\mathbf{u}\notag \\
&+\lambda_R n_I (\partial_t+\mathbf{u} \cdot \nabla -\nu \nabla^2)\mathbf{u}=v^2 n \mathbf{E}.
\end{align}
This is the second main result of the paper. Compared to the case of pristine graphene without RSO coupling~\cite{Narozhny2019}, one finds a new term on the left-hand side of the equation. Since the spin-orbit coupling gives rise to an additional term in the enthalpy~(\ref{Macroscopic equations for densities and currents}), it affects the time derivative and the convective term.

\section{Conclusions}
\label{Sec:V}
To summarize, we have derived the hydrodynamic Navier-Stokes equation for two-dimensional graphene-like materials in the presence of RSO coupling. Compared to the result without spin-orbit coupling, the RSO interaction modifies the viscosity and gives rise to an additional term in the Navier-Stokes equation, which is similar to the convective term. The reason for this is the modification of the spectrum of the system due to the presence of the RSO interaction, which results in the addition of a term $\lambda_R n_I$ to the enthalpy. As the momentum continuity equation is sensitive to the explicit form of the spectrum, this influences the final form of the hydrodynamic Navier-Stokes equation. In addition, we have derived the two-particle scattering rate explicitly in the presence of RSO coupling. This has allowed us to derive the corrections to the effective kinematic viscosity resulting from the RSO coupling.

\acknowledgments
We are grateful to Boris Narozhny and Kristof Moors for fruitful discussions. The authors acknowledge financial support from the National Research Fund Luxembourg under Grants C19/MS/13579612/HYBMES, C21/MS/15752388/NavSQM, PRIDE19/14063202/AC\-TIVE and INTER/MOBILITY/2022/MS/17549827/AndMTI.

\appendix
\begin{widetext}
\section{Calculation of matrix element \texorpdfstring{$\langle n^{\prime} \vk^{\prime}|e^{i\vq \vr}|n \vk\rangle$}{}}
\label{Sec:A}
In this Appendix we calculate the form of the matrix element $\langle n^{\prime} \vk^{\prime}|e^{i\vq \vr}|n \vk\rangle$~\cite{Shung1986}. First we demonstrate the calculations for pristine graphene and then generalize the results for RSO interactions. The orthonormal eigenvectors of Hamiltonian~(\ref{Pristine graphene's Hamiltonian}) have the form
\begin{equation}
\label{Sec:A. Eigenectors of pristine graphene}
\psi^{n}_{\vk}(\vr)=\frac{1}{\sqrt{2}}\left[\mathcal{U}_{\vk,A}(\vr)-n\frac{g^{\ast}(\vk)}{|g(\vk)|}\mathcal{U}_{\vk,B}(\vr)\right], \quad \mathcal{U}_{\vk,j}(\vr)=\frac{1}{\sqrt{N}}\sum_{\mathbf{R}_{j^{\prime}}}e^{i\vk(\mathbf{R}_{j^{\prime}}+\boldsymbol{\tau}_{j})}\phi_z(\vr-\mathbf{R}_{j^{\prime}}-\boldsymbol{\tau}_j), \quad j=A,B,	
\end{equation}
where $n=+1$ ($n=-1$) corresponds to the conduction (valence) band, the subscripts $A$ and $B$ correspond to the two atoms in the unit cell, the pre-factor is a complex number $g(\vk)=k_x+ik_y=|\vk|e^{i \theta_{\vk}}$, where $\theta_{\vk}$ defines the direction of the 2D vector $\vk$ and $N$ is a normalization coefficient (number of unit cells). The localized function $\phi_z(\vr)$ corresponds to a $p_z$ orbital of atoms $A$ and $B$, $\mathbf{R}_{j^{\prime}}$ is a lattice vector, and $\boldsymbol{\tau}_j$ denotes the position of the atom $j \in \{A,B\}$ within the unit cell. For instance, one can set $\boldsymbol{\tau}_A=0$, in which case $\boldsymbol{\tau}_B$ denotes the vector from atom $A$ to $B$. The $p_z$ orbitals are approximated by hydrogen wave functions with an effective core charge $Z \simeq 3.2$ and Bohr radius $a_0=5.3$\AA~\cite{Zener}, namely
\begin{equation}
\label{Sec:A. Hydrogen wave function}
\phi_z(\vr)=\phi_{n=2,l=1,m=0}(r,\theta,\varphi)=\frac{Z^{3/2}}{4\sqrt{2\pi}a^{3/2}_0}\frac{Zr}{a_0}e^{-\frac{Zr}{a_0}}\cos\theta.
\end{equation}
To calculate the matrix element $F^{nn^{\prime}}_{\vk\vk^{\prime}}$ we insert the projection operator $\int d\vr |\vr\rangle \langle \vr|=1$, so that we have
\begin{equation}
\label{Sec:A. Inserting of projection operator into matrix element}
F^{nn^{\prime}}_{\vk\vk^{\prime}}=\int d\vr \langle n^{\prime} \mathbf{k^{\prime}}|\vr\rangle e^{i\vq \vr} \langle \vr| n\vk\rangle=\int d\vr \psi^{n^{\prime}} _{\vk^{\prime}}(\vr)e^{i\vq \vr}\psi^n_{\vk}(\vr)=\frac{1}{2}\left(1+nn^{\prime}\frac{g^{\ast}(\vk)g(\vk^{\prime})}{|g^{\ast}(\vk)||g(\vk^{\prime})|}\right) I(\vq;\vk,\mathbf{k^{\prime}}),
\end{equation}
where it is assumed that the orbitals of the atoms $A$ and $B$ are orthogonal, and the auto-correlation term is given by the following integral expression
\begin{equation}
\label{Sec:A. Auto correlation formula}
I(\vq;\vk,\mathbf{k^{\prime}})=\int d\vr \mathcal{U}^{\ast}_{\vk^{\prime},A}(\vr)e^{i\vq \vr}\mathcal{U}_{\vk,A}(\vr)=\int d\vr \mathcal{U}^{\ast}_{\vk^{\prime},B}(\vr)e^{i\vq \vr}\mathcal{U}_{\vk,B}(\vr)=\frac{1}{N} \sum_{\mathbf{R}_j}e^{i(\vk+\vq-\mathbf{k^{\prime}})\mathbf{R}_j} \mathcal{I}(\vq),
\end{equation}
where the inner integral is denoted by $\mathcal{I}(\vq)=\int d\vr \sum_{\boldsymbol{\delta}} e^{i\vk \boldsymbol{\delta}}\phi_z(\vr)e^{i\vq \vr}\phi_z(\vr+\boldsymbol{\delta})$ with $\boldsymbol{\delta}=\mathbf{R}_{j^{\prime}}-\mathbf{R}_j$. The main contribution to this integral comes from $\boldsymbol{\delta}=0$, so we approximate the integral as follows
\begin{equation}
\label{Sec:A. Integral approximation with zero delta}
	\mathcal{I}(\vq) \simeq \int d\vr \phi_z(\vr)e^{i\vq \vr}\phi_z(\vr)=\frac{1}{(1+|\mathbf{Q}|^2)^3}-\frac{6 |\mathbf{Q}|^2}{(1+|\mathbf{Q}^2)|^4}, \quad |\mathbf{Q}|=|\vq|a_0/Z.
\end{equation}
At small momentum transfer $|\mathbf{Q}| \ll 1$, the expression~(\ref{Sec:A. Integral approximation with zero delta}) is of the order of unity, and taking into account the conservation of momentum in the scattering process, i.e., $(1/N)\sum_{\mathbf{R}_j}\exp[i(\vk+\vq-\mathbf{k^{\prime}})\mathbf{R}_j]=1$, we arrive at the final result for matrix element in Eq.~(\ref{Sec:A. Inserting of projection operator into matrix element}), namely
\begin{equation}
\label{Sec:A. Dirac factor without spin-orbit coupling}
F^{nn^{\prime}}_{\vk \mathbf{k^{\prime}}}=\frac{1}{2}\left(1+nn^{\prime}\frac{g^{\ast}(\vk)g(\vk^{\prime})}{|g^{\ast}(\vk)||g(\vk^{\prime})|}\right).
\end{equation}
In presence of RSO interaction the orthonormal eigenvectors of Eq.~(\ref{Zero Hamiltonian in second quantization}) are given by
\begin{equation}
\label{Sec:A. Eigenvector of Hamiltonain with spin-orbit coupling}
\psi^n_{\vk}(\vr)=\frac{1}{\sqrt{2}}\xi^{n}_{\vk}\left[\mathcal{U}_{\vk,A\uparrow}(\vr)+ng^{n}_{\mathbf{k,1}}\mathcal{U}_{\vk,A\downarrow}(\vr)+g^n_{\vk,2}\mathcal{U}_{\vk,B\uparrow}(\vr)+ng^{n}_{\mathbf{k,3}}\mathcal{U}_{\vk,B\downarrow}(\vr)\right],
\end{equation}
where $\xi^{n}_{\vk} = \sqrt{1+(\zeta^n_{\vk})^2}$ and $\zeta^n_{\vk} = \epsilon_{n\vk}/(v|\vk|)$. Further ignoring the cross correlations and repeating the same steps as in case of pristine graphene above, we arrive at Eq.~(\ref{Dirac factor in presence of spin-orbit coupling}) of the main text.

\section{Derivation of the macroscopic Eqs.~(\ref{Macroscopic equations for densities and currents})}
\label{Sec:B}
In this Appendix we derive the expressions which relate the macroscopic densities and currents using the local equilibrium distribution function~(\ref{Local equilibrium distribution function}) and the definitions of particle, heat, momentum densities and currents. In our derivations we omit a degeneracy factor $\mathcal{N}=2$ due to the valley degeneracy. First, we demonstrate that $\mathbf{j}(\vr,t)=n(\vr,t) \mathbf{u}(\vr,t)$ by considering the following difference
\begin{equation}
\label{Derivation of j=nu}	
\mathbf{j}-n\mathbf{u}=\sum_n \int \frac{d\vk}{(2\pi)^2} f_{n\vk}\partial_{\vk}(\epsilon_{n\vk}-\mathbf{u} \cdot \vk-\mu)
=\frac{1}{\beta}\sum_n \int \frac{d\vk}{(2\pi)^2} \partial_{\vk}\left[\log\left(1+e^{-j[\epsilon_{n\vk}-\mathbf{u}\cdot \vk-\mu]}\right)\right]=0,
\end{equation}
where we used that $\epsilon_{n\vk}=\mp \lambda_R+\sqrt{v^2\vk^2+\lambda^2_R} > \mathbf{u} \cdot \vk$, since the Fermi velocity is larger then the drift velocity, $v>|\mathbf{u}|$. An analogous calculation leads to $\mathbf{j}_I(\vr,t)=n_I(\vr,t) \mathbf{u}(\vr,t)$. Next, we turn to the energy current where we need to prove that $\mathbf{j}_E=W \mathbf{u}$. Since $W=n_E+P$ this is equivalent to $\mathbf{j}_E-n_E \mathbf{u}=P\mathbf{u}$. This is shown by straightforward calculations

\begin{align}
\label{Derivation of jE=Wu}
&
    \mathbf{j}_E-n_E \mathbf{u}
=
    \sum_n \int \frac{d\vk}{(2\pi)^2} \epsilon_{n\vk} f_{n\vk} \partial_{\vk} (\epsilon_{n\vk}-\mathbf{u} \cdot \vk-\mu)
=
    -\frac{1}{\beta} \sum_n \int \frac{d\vk}{(2\pi)^2} \epsilon_{n\vk} \partial_{\vk} \log\left[ 1+e^{-\beta(\epsilon_{n\vk}-\mathbf{u}\cdot \vk-\mu)}\right] \\
&
=
    \frac{1}{\beta} \sum_n \int \frac{d \vk}{(2\pi)^2} \log\left[1+e^{-\beta(\epsilon_{n \vk}-\mathbf{u}\vk-\mu)}\right] \partial_{\vk} (\epsilon_{n \vk}-\mathbf{u}\vk+\mathbf{u}\vk-\mu) \notag \\
&=
    \underbrace{
        \frac{\mathbf{u}}{\beta} \sum_n \int \frac{d \vk}{(2\pi)^2} \log\left[1+e^{-\beta(\epsilon_{n \vk}-\mathbf{u}\vk-\mu)}\right]
    }_{=\mathbf{u}P}
+
    \underbrace{\frac{1}{\beta} \sum_n \int \frac{d \vk}{(2\pi)^2} \partial_{\vk} (\epsilon_{n \vk}-\mathbf{u}\vk-\mu) \log\left[1+e^{-\beta (\epsilon_{n \vk}-\mathbf{u}\vk-\mu)}\right]
    }_{=0}=P\mathbf{u}, \notag
\end{align}
where the definition of pressure in Eq.~(\ref{Pressure}) was used. Thus we obtained the third expression of Eqs.~(\ref{Macroscopic equations for densities and currents}) in the main text. Next, we derive the expression for the momentum density, which can be presented as a product of enthalpy, imbalance density, and drift velocity. Let us consider the $x$ component of the energy current
\begin{equation}
\label{Derivation of nk=tilde Wu/v2}
\begin{aligned}
    j^x_{E}
&=
    \sum_n \int \frac{d \vk}{(2\pi)^2} \epsilon_{n \vk} \frac{v^2 k_x}{v^2 \vk^2+\lambda^2_R}f_{n \vk}
=
    \sum_n \int \frac{d \vk}{(2\pi)^2} \left(1-\frac{n \lambda_R}{\sqrt{v^2 \vk^2+\lambda^2_R}}\right) v^2 k_x f_{n \vk} \notag \\
&=
    v^2 \underbrace{\sum_n \int \frac{d\vk}{(2\pi)^2} k_x f_{n \vk}}_{n^x_{\vk}}
-
    \lambda_R \underbrace{\sum_n n \int \frac{d\vk}{(2\pi)^2} \frac{v^2 k_x}{\sqrt{v^2 \vk^2+\lambda^2_R}} \epsilon_{n \vk} f_{n \vk}}_{j^x_{I}}
=
    v^2 n^x_{\vk}-\lambda_R j^x_{I}.
\end{aligned}	
\end{equation}
and analogously for the $y$ component. Therefore, we get the fourth relation of Eqs.~(\ref{Macroscopic equations for densities and currents}) of main text, i.e.,
\begin{equation}
\label{Derivation of nk=tilde Wu/v2. The final expression}
\mathbf{n}_{\vk}=v^{-2} \mathbf{j}_E+v^{-2} \lambda_R \mathbf{j}_I=v^{-2} \left(W+\lambda_R n_I \right) \mathbf{u}.
\end{equation}
Finally, we derive the last relation of Eqs.~(\ref{Macroscopic equations for densities and currents}) for the stress tensor. By definition, the diagonal element $\Pi^{xx}_E$ of the stress tensor is given by
\begin{align}
    \Pi^{xx}_E
&=
    \sum_n \int \frac{d\vk}{(2\pi)^2} k_x v^x_{n \vk}f_{n \vk}
=
    \sum_n \int \frac{d\vk}{(2\pi)^2} k_x \partial_{k_x}(\epsilon_{n \vk}-u_x k_x-u_y k_y -\mu+u_x k_x) f_{n \vk} \notag \\
&=
    u_x \underbrace{ \sum_n \int \frac{d\vk}{(2\pi)^2} k_x f_{n \vk}}_{n^x_{\vk}}
-
    \frac{1}{\beta} \sum_n \int \frac{d \vk}{(2\pi)^2} k_x \partial_{k_x} \log\left[1+e^{-\beta(\epsilon_{n \vk}-u_x k_x-u_y k_y-\mu)}\right] \notag \\
&=
    \underbrace{u_x n^x_{\vk}}_{v^{-2} \tilde{W} u_x u_x}
    +\underbrace{\frac{1}{\beta} \sum_n \int \frac{d\vk}{(2\pi)^2} \log\left[1+e^{-\beta(\epsilon_{n \vk}-u_x k_x-u_y k_y-\mu)}\right]}_{P}
    =P+v^{-2} \tilde{W} u_x u_x,
\end{align}
where $\tilde{W}=W+\lambda_R n_I$. Moreover, the off-diagonal component $\Pi^{xy}_{E}$ of the stress tensor is given by
\begin{align}	
\label{Derivation of off-diagonal stress tensors component}
    \Pi^{xy}_{E}
&=
    \sum_n \int \frac{d\vk}{(2\pi)^2} k_x v^y_{n \vk} f_{n \vk}
=
    \sum_n \int \frac{d\vk}{(2\pi)^2} k_x \partial_{k_y}(\epsilon_{n \vk}-u_y k_y-u_x k_x -\mu+u_y k_y) f_{n \vk} \notag \\
&=
    u_y \underbrace{\sum_n \int \frac{d\vk}{(2\pi)^2} k_x f_{n \vk}}_{n^x_{\vk}}
-
    \frac{1}{\beta} \sum_n \int \frac{dk_x}{2\pi} k_x \underbrace{ \int \frac{dk_y}{2\pi} \partial_{k_y} \log[1+e^{-\beta(\epsilon_{n \vk}-u_x k_x-u_y k_y-\mu)}]}_{=0} \notag \\
&=
    n^x_{\vk} u_y
=
    v^{-2}[W+\lambda_R n_I] u_x u_y.
\end{align}
and similar expressions can be obtained for $\Pi^{yy}_E$ and $\Pi^{yx}_E$. In summary, this leads to the last lines in Eqs.~(\ref{Macroscopic equations for densities and currents}) of the main text.

 \section{Effects of spin-orbit coupling on electron-electron scattering rate and viscosity} \label{sec:collision}
 We consider the Boltzmann equation including only the electron-electron scattering integral,
 \begin{equation}
\frac{\partial f_{n_1\vk_1}}{\partial t}+\frac{\partial \epsilon_{n_1\vk_1}}{\partial \vk_1}\frac{\partial f_{n_1 \vk_1}}{\partial \vr}-\frac{\partial \varphi}{\partial \vr} \frac{\partial f_{n_1 \vk_1}}{\partial \vk_1}=\mathcal{I}[f_{n_1\vk_1}],\label{eq:kineticd}
 \end{equation}
where the scattering integral is given by
\begin{equation}
\mathcal{I}[f_1] =-2\pi \mathcal{N}\int_{234} \delta_{1234}W_{12}^{34} \lb f_1 f_2(1-f_{3})  (1- f_4) - (1-f_1) (1- f_2)  f_{3}  f_4 \rb , \label{eq:s}
\end{equation}
where $\mathcal{N}=2$ accounting valley degeneracy.
We have shortened the indices for simplicity as $\{1,2,3,4\}=\{n_1\vec {k}_1,n_2\vec {k}_2,n_3(\vec{k}_1+\vec{q}), n_4(\vec{k}_2-\vec{q})\}$ and $\delta_{1234}$ contains the Dirac delta functions for both energy and momentum conservation. We simplify the band index degrees of freedom by neglecting interband transitions ($n_1=n_3$) and interband scattering ($n_1=n_2$). The interband transition normally occurs on fast time scales $\omega \gg 1/\tau_{ee}$ relevant, e.g., to optics which is not amenable for hydrodynamics. Interband scattering between the lower and upper conduction band will give mutual drags that cancel each other in the Navier-Stokes equation [summing up two $c$ bands contribution in Eq.~\eqref{eq:kineticd}]. To simplify the calculation of scattering integral, we assume a small drift velocity so that the local equilibrium distribution is given by,
\begin{equation}
f_{n,\vec {k}}^\zero = \frac{1}{1+ \exp[\beta (\eps_\vec {k}^n-\mu )]}
\end{equation}
We linearize the collision integral~\eqref{eq:s} by writing
\begin{eqnarray}
f_{n,\vec {k}} &=& f^{\zero}_{n,\vec {k}} +\delta f =f^{\zero}_{n,\vec {k}}   -  \frac{\partial f^{\zero}_\vec {k}  }{\partial \eps} F(\vec r,\theta_\vec {k})
\end{eqnarray}
where the nonequilibrium distribution $\delta f$ is assumed to be valid at low temperatures where $- \partial_\eps f^{\zero}_\vec {k} $ can be approximates as a delta function peaked at $\mu$ and the $\vec {k}$ dependence on $F$ only depends on the azimuthal angle $\theta_\vec {k}$ between $\vec {k}$ and its component $k_x$.  We further expand the nonequilibrium part of distribution into angular harmonics:
\begin{equation}
F(\vec r,\theta_\vec {k})=\sum_{n=-\infty}^\infty e^{in \theta_\vec {k}} \ft F_{n} (\vec r) \label{eq:angular}
\end{equation}
We see that $\ft F_0$ is related to the density fluctuations
\begin{eqnarray}
n (\vec r,t) &=& \frac{1}{(2\pi)^2}\sum_{m} \int d^2\vec {k} \int d \eps \delta (\eps-\eps_{\vec k}^m) \lp f(\vec r,\vec {k},t)-f^\zero (\eps)\rp
=  g_F \ft F_0 (\vec r,t),\label{eq:dens}
\end{eqnarray}
where $ g_F = \frac{1}{2\pi} \sum_m \int k dk \delta (\mu - \eps_\vec k^m)$ is the local density of states at the Fermi level.
Moreover, the functions $\ft F_{\pm 1}$ are related to the current density,
\begin{eqnarray}
\vec j (\vec r,t) &=& \frac{1}{(2\pi)^2} \sum_m \int d^2 \vec {k}\ \vec v_\vec {k}^m \lp f(\vec r,\vec k,t)-f^\zero(\eps)  \rp,\nn
&=& \frac{1}{2}  v_F \begin {pmatrix}
\ft F_1 (\vec r,t) + \ft F_{-1} (\vec r,t)\\
i \left[ \ft F_1 (\vec r,t) - \ft F_{-1} (\vec r,t)\right]
\end {pmatrix}\equiv\bar n \vec u(\vec r,t), \label{eq:current}
\end{eqnarray}
where $\bar n = g_F\mu$ is the equilibrium density, $v_F=\frac{1}{2\pi}\sum_m \int k dk \delta (\mu - \eps_\vec k^m) v_{\vec k}^m$ and this equation defines the drift velocity $\vec u(\vr, t)$.
The functions $\ft F_{\pm 2}$ are related to the stress tensor,
\begin{eqnarray}
{\Pi}^{i,j} &=&\frac{1}{(2\pi)^2}\sum_m \int d^2 \vec {k}\ k_i  v_{k,j}^m \int d\eps \delta(\eps-\eps_\vec k^m) (f(\vec r,t) -f^\zero(\eps)),\nn
{ \Pi}^{x,x}
&=& \frac{\pi_F}{4}\lp 2\ft F_0 (\vec r,t) + \ft F_2(\vec r,t) + \ft F_{-2}(\vec r,t)\rp,\nn
\Pi^{x,y} = \Pi^{y,x}&=& \frac{\pi_F}{4i} \lp \ft F_{-2}(\vec r,t) -\ft F_{2}(\vec r,t)\rp,  \\
\Pi^{y,y} &=& \frac{\pi_F}{4}\lp 2\ft F_0 (\vec r,t) - \ft F_2(\vec r,t) - \ft F_{-2}(\vec r,t)\rp,\nonumber\end{eqnarray}
where $\pi_F =\frac{1}{2\pi} \sum_m \int k dk \delta(\mu-\eps_\vec k^m) k v_\vec k ^m$.
We can express this tensor in a compact form as
\begin{equation}
{\boldsymbol \Pi} = \frac{\pi_F}{4} \lb 2\ft F_0 + (\ft F_2 +\ft F_{-2})\tau_z+ i(\ft F_2-F_{-2})\tau_x\rb,\label{eq:stress}
\end{equation}
where $\tau_{x,z}$ are the Pauli matrices.

Next, we multiply Eq.~\eqref{eq:kineticd} by $e^{-im \theta_\vec {k}}$ and integrate over $\vec {k}$ to obtain
\begin{eqnarray}
g_F \partial_t \ft F_m (\vec r,t)+\frac{v_F}{2} \lb \partial_x (\ft F_{m-1}+\ft F_{m+1})-i\partial_y (\ft F_{m-1}-\ft F_{m+1})\rb&&\nn
-\frac{ v_F}{2}\lb E_x(\delta_{m,1}+\delta_{m,-1})-iE_y(\delta_{m,1}-\delta_{m,-1})\rb
&=& -\gamma_m \ft F_m \label{eq:recursive}
\end{eqnarray}
It will be obvious later that we can obtain the right hand side from linearization of the scattering integral and only even harmonics $|m|>1$ give nonzero $\gamma_m$.
For $m=0$, we obtain the continuity equation
\begin{equation}
\partial_t \bar n+\nabla \vec j =0.
\end{equation}
For $m=\pm 1$, we obtain the linearized Navier-Stokes equation
\begin{equation}
\partial_t \vec u +\frac{1}{\rho} \nabla \cdot {\boldsymbol \Pi} -\frac{ 1}{\bar m} \vec E =0,\label{eq:NSE}
\end{equation}
where $\bar m=\bar n g_F/v_F^2$ is proportional to mass,  $\rho=\bar n g_F \pi_F/v_F^2 $ is the mass density and in graphene $\rho = \tilde W/v^2$.
We can approximately close the recursion relation~\eqref{eq:recursive} by setting $\ft F_m=0$ for $|m|\ge 3$~\cite{pellegrino17-hallvisc,Lucas2018}. For $m=2$ and considering components at frequency $\omega$, i.e., $\partial_t \ft F_m=-i\omega \ft F_m$, we obtain
\begin{eqnarray}
\frac{v_F}{2} \lb \partial_x \ft F_{1}-i\partial_y \ft F_{1}\rb &=&-\lp \gamma_2-i\omega g_F \rp \ft F_2\\
\frac{v_F}{2} \lb \partial_x \ft F_{-1}+i\partial_y \ft F_{-1}\rb &=&-\lp \gamma_{-2} -i\omega g_F \rp \ft F_{-2}.\label{eq:F2}
\end{eqnarray}
We will see later that $\gamma_2=\gamma_{-2}$. Using the relationship in Eq.~\eqref{eq:F2}, the stress tensor in Eq.~\eqref{eq:stress} becomes
\begin{eqnarray}
{\boldsymbol \Pi} &=& P - \rho \nu_\omega \lb  (\partial_x u_x-\partial_y u_y)\tau_z- (\partial_x u_y +\partial_y u_x)\tau_x\rb,\label{eq:stress2app}
\end{eqnarray}
and the Navier-Stokes equation becomes
\begin{equation}
-i\omega \vec u + \nabla P -\nu_\omega \nabla^2 \vec u - \frac{\bar n}{\rho} \vec E =0,
\end{equation}
where $ P=  n (\vec r,t) \pi_F/2  $ is the pressure,  and the kinematic viscosity is given by
\begin{equation}
\nu_\omega=\frac{v_F^2}{4\displaystyle g_F \lp \gamma_2-i\omega g_F \rp }.
\end{equation}
The stress tensor in Eq.~\eqref{eq:stress2app} does not contain the convective term because we focus on linear term but it captures the dissipative term induced by relaxation of the second harmonics $\ft F_{\pm 2}$.

Next, we discuss the linear collision integral. Linearizing the collision integral means retaining the linear order of the distribution function products,
\begin{eqnarray}
f_1 f_2\lp1-f_{3} \rp \lp 1- f_4\rp &=&   (f_1^\zero +\delta f_1)( f_2^\zero+\delta f_2)\lp 1-(f_{3}^\zero+\delta f_{3}) \rp \lp 1- (f_4^\zero +\delta f_4)\rp\nn
&\approx& f_1^\zero f_2^\zero\lp1-f_{3}^\zero \rp \lp 1- f_4^\zero\rp + \delta f_1 f_2^\zero\lp1-f_{3}^\zero \rp \lp 1- f_4^\zero\rp\nn
&& + f_1^\zero \delta f_2\lp1-f_{3}^\zero \rp \lp 1- f_4^\zero\rp  + f_1^\zero  f_2^\zero(-\delta f_{3}) \lp 1- f_4^\zero\rp\nn
&& + f_1^\zero  f_2^\zero\lp1-f_{3}^\zero \rp (-\delta f_4^\zero).
\end{eqnarray}
Analogously, we obtain
\begin{eqnarray}
(1-f_1) (1-f_2) f_{3}  f_4 &\approx& (1-f_1^\zero)(1- f_2^\zero) f_{3}^\zero  f_4^\zero +(-\delta f_1) (1-f_2^\zero)f_{3}^\zero f_4^\zero\nn
&& + (1- f_1^\zero) (-\delta f_2) f_3^\zero  f_4^\zero  + (1-f_1^\zero)(1-  f_2^\zero)(\delta f_3) f_4^\zero\nn
&& + (1-f_1^\zero)(1-  f_2^\zero)f_3^\zero  (\delta f_4^\zero).
\end{eqnarray}
Therefore,
\begin{eqnarray} f_1 f_2\lp1-f_3 \rp \lp 1- f_4\rp - (1-f_1) (1-f_2) f_3  f_4
&\approx&
\delta f_1 \lb f_2^\zero \lp 1- f_3^\zero \rp \lp 1-f_4^\zero\rp + (1-f_2^\zero) f_3^\zero f_4^\zero\rb\nn
&&+ \delta f_2 \lb f_1^\zero \lp 1- f_3^\zero \rp \lp 1-f_4^\zero\rp + (1-f_1^\zero) f_3^\zero f_4^\zero\rb \nn
&&-\delta f_3 \lb f_1^\zero f_2^\zero \lp 1-f_4^\zero\rp + \lp 1-f_1^\zero\rp \lp 1-f_2^\zero \rp f_4^\zero    \rb \nn
&&- \delta f_4 \lb f_1^\zero f_2^\zero \lp 1-f_3^\zero\rp + \lp 1-f_1^\zero\rp \lp 1-f_2^\zero \rp f_3^\zero    \rb ,\nn
&\approx&
f_1^\zero f_2^\zero\lp1-f_3^\zero \rp \lp 1- f_4^\zero\rp \lb -h_1 -h_2 + h_3 + h_4\rb,
\end{eqnarray}
where we have used $  f_1^\zero f_2^\zero(1-f_3^\zero )( 1- f_4^\zero) - (1-f_1^\zero) (1-f_2^\zero) f_3^\zero  f_4^\zero=0$ as a consequence of energy and number conservation, and $\delta f_1 = -h_1 f_1^\zero (1-f_1^\zero) $, where $h_1 = \beta F(\vec r, \theta_1)$.
The linearized collision integral reads
\begin{equation}
\mathcal{I}[f_1] =4\pi\int_{234} \delta_{1234}W_{12}^{34}  f_1^\zero f_2^\zero(1-f_3^\zero)  (1- f_4^\zero) h_{12}^{34}, \label{eq:se-lin}
\end{equation}
where $h_{12}^{34} = (h_1+h_2) -(h_3+h_4)$. Now we substitute the harmonic expansion into $h_i$, we can write down the $n$th eigenmode as
\begin{equation}
\mathcal{I}_n[f_1] =4\pi\beta \ft F_n(\vec r) \int_{234} \delta_{1234}W_{12}^{34}  f_1^\zero f_2^\zero(1-f_3^\zero)  (1- f_4^\zero) (e^{in\theta_1}+e^{in\theta_2}-e^{in\theta_3}-e^{in\theta_4}), \label{eq:se-lin_n}
\end{equation}
Thus, this leads to the following definition of the electron-electron scattering rate
\begin{equation}
\gamma_n =4\pi\beta  \int_{1234} \delta_{1234}W_{12}^{34}  f_1^\zero f_2^\zero(1-f_3^\zero)  (1- f_4^\zero) (1+e^{in(\theta_2-\theta_1)}-e^{in(\theta_3-\theta_1)}-e^{in(\theta_4-\theta_1)}). \label{eq:ee-rate}
\end{equation}
where the last factor in Eq.~\eqref{eq:ee-rate} comes from taking the integral $\int_1 e^{-in\theta_1} \mathcal{I}[f_1]$.
\begin{figure}
\centering
\includegraphics[width=6cm]{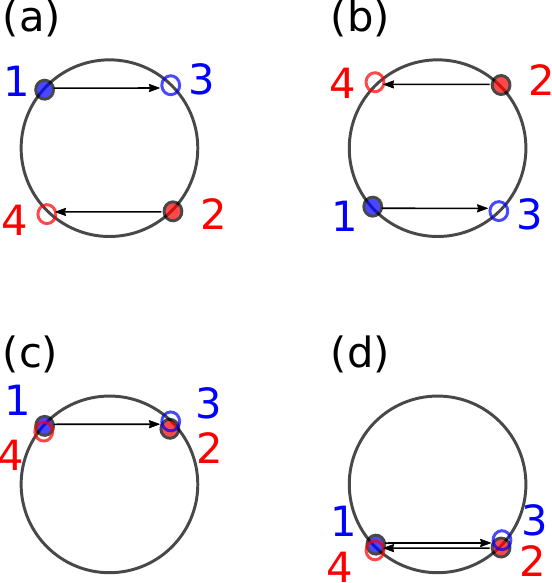}
\caption{Possible states for e-e collision. (a) and (b) are head-on collisions while (c) and (d) are exchange processes. }
\label{fig:ee}
\end{figure}

We schematically look at the possible electron-electron (e-e) collisions in Fig.~\ref{fig:ee}~\cite{ledwidth19}. In Fig.~\ref{fig:ee}, we consider a momentum transfer $\vec q$ along the $k_x$ axis. We can do so without loss of generality because we can rotate Fig.~\ref{fig:ee} around an angle $\theta_\vec q$ and the resulting e-e collision rate is invariant. The kinematic restrictions due to momentum and energy conservations leave only a few possibilities for scattering processes. The first type is a head-on collision with  $\vec {k}_2=-\vec {k}_1$ or $\theta_2=\pi + \theta_1$, $\theta_3=-\theta_1$ and $\theta_4=\pi-\theta_1$ as depicted in Figs.~\ref{fig:ee}(a) and (b). The second type is an exchange process where $\vec {k}_1=\vec {k}_4$ and $\vec {k}_2=\vec {k}_3$. Using this condition in Eq.~\eqref{eq:se-lin_n}, we find that the exchange process gives $\mathcal{I}_n[f_1]=0$ and thus does not relax the e-e scattering. Focussing on the head-on process, we note that $n=0$ and $n=1$ also give $\mathcal{I}_n[f_1]=0$. This is expected as a result of particle number, energy ($n=0$) and momentum ($n=1$) conservation. Further calculations show that all odd $n$ modes give zero $\mathcal{I}_n[f_1]$.
For even $n$, we obtain
\begin{eqnarray}
e^{-in\theta_1}h_{12}^{34}(n) &=& (1+e^{in\pi}-e^{in(-2\theta_1)}-e^{in(\pi-2\theta_1)})+\text{c.c.}\nn
&=& 4(1-\cos(2n\theta_1)),\quad \text{where}\quad \sin \theta_1= \frac{q}{2k_F}
\end{eqnarray}
For $n=2$, we obtain
\begin{equation}
e^{-in\theta_1}h_{12}^{34}(2)=32\lb \lp \frac{q}{2k_F}\rp^2 - \lp \frac{q}{2k_F}\rp^4\rb
\end{equation}

Next,  we will consider the actual e-e interaction matrix element $W_{12}^{34}=|V_{12}^{34}|^2= |V_q \braket{3|1} \braket{4|2}|^2$, neglecting the interference term of Eq.~\eqref{Transition matrix element}. For the $4\times 4$ Hamitonian of graphene with SOC we use the corresponding eigenstate,
\begin{equation}
|\vec k\rangle = \frac{1}{\sqrt 2}\frac{e^{-i\theta_{\vec k}}}{\sqrt{1+(\eps_\vec k/vk)^2}}
\lp
 i e^{-i\theta_\vec k} ,\quad
-i \frac{\eps_\vec k}{v k},\quad
\frac{\eps_\vec k}{vk},\quad
e^{i\theta_\vec k}
\rp^T.
\end{equation}
For different bands, the eigenstates differ by energy dispersion. For simplicity, we can  suppress the band index  unless we explicitly write the dispersion, i.e. $\eps_\vec k^\pm = \pm \lambda_R +\sqrt{\lambda_R^2+(v k)^2}$, where $-$ for $c_1$ and $+$ for $c_2$.

We obtain
\begin{eqnarray}
|\langle {\vec{k}_1+\vec{q}}|{\vec{k}_1}\rangle|^2 &=&\frac{v^2 k_1|{\vec{k}_1}+\vec q| \cos(\theta_{\vec{k}_1}-\theta_\vec{k_1+q})+\eps_{\vec{k}_1}\eps_\vec{k_1+q}}{v^2k_1|{\vec{k}_1+\vec{q}}|\sqrt{1+(\eps_{\vec{k}_1}/vk_1)^2} \sqrt{1+(\eps_{\vec{k}_1+\vec{q}}/v|{\vec{k}_1+\vec{q}}|)^2}}\nn
|\langle {\vec{k}_2-\vec{q}}|\vec k_{2}\rangle|^2 &=&\frac{v^2 k_2|{\vec{k}_2}-\vec q| \cos(\theta_{\vec{k}_2}-\theta_\vec{k_2-q})+\eps_{\vec{k}_2}\eps_\vec{k_2-q}}{v^2k_2|{\vec{k}_2-\vec{q}}|\sqrt{1+(\eps_{\vec{k}_2}/vk_2)^2} \sqrt{1+(\eps_{\vec{k}_2-\vec{q}}/v|{\vec{k}_2-\vec{q}}|)^2}}.\nonumber
\end{eqnarray}
We need to express $\cos(\theta_{{\vec{k}_1+\vec{q}}}-\theta_{\vec{k}_1})$ in terms of $\phi=\theta_\vec q-\theta_{\vec{k}_1}$,
\begin{eqnarray}
\cos (\theta_{{\vec{k}_1+\vec{q}}}-\theta_{\vec{k}_1}) &=& \cos \theta_{{\vec{k}_1+\vec{q}}}\cos \theta_{{\vec{k}_1}} +\sin \theta_{\vec{k}_1+\vec{q}} \sin \theta_{\vec{k}_1} \nn
&=& \frac{(\vec{k_1+q})_x}{|{\vec{k}_1+\vec{q}}|} \cos \theta_{\vec{k}_1} + \frac{(\vec{k_1+q})_y}{|{\vec{k}_1+\vec{q}}|}  \sin \theta_\vec p \nn
&=& \frac{k\cos\theta_{\vec{k}_1} + q \cos \theta_\vec q}{|\vec{k_1+q}|} \cos\theta_{\vec{k}_1} + \frac{k_1\sin\theta_{\vec{k}_1} + q \sin \theta_\vec q}{|\vec{k_1+q}|} \sin\theta_{\vec{k}_1}   \nn
&=& \frac{k_1 + q(\cos \theta_\vec q \cos \theta_{\vec{k}_1} + \sin \theta_\vec q \sin \theta_{\vec{k}_1})}{|{\vec{k}_1+\vec{q}}|}\nn
&=& \frac{k_1 + q\cos\phi}{|{\vec{k}_1+\vec{q}}|}
\end{eqnarray}
In the same way, we obtain  $\cos(\theta_\vec{k_2-q}-\theta_{\vec{k}_2})=(k_2-q \cos \phi_2)/ |\vec {k_2 -q}|$, where $\phi_2=\theta_\vec q-\theta_\vec{ k_2}$. As a result, we have
\begin{eqnarray}
|\langle {\vec{k}_1+\vec{q}}|{\vec{k}_1}\rangle|^2 &=&\frac{v^2 k_1 (k_1+q \cos\phi)+\eps_{\vec{k}_1}\eps_\vec{k_1+q}}{v^2k_1|\vec {k_1 + q}|\sqrt{1+(\eps_{\vec{k}_1}/vp)^2} \sqrt{1+(\eps_{\vec{k}_1+\vec{q}}/v|{\vec{k}_1+\vec{q}}|)^2}}\nn
|\langle {\vec{k}_2-\vec{q}}|\vec k_{2}\rangle|^2 &=&\frac{v^2 k_2 (k_2-q \cos\phi_2)+\eps_{\vec{k}_2}\eps_\vec{k_2-q}}{v^2k_2|{\vec{k}_2-\vec{q}}|\sqrt{1+(\eps_{\vec{k}_2}/vk_2)^2} \sqrt{1+(\eps_{\vec{k}_2-\vec{q}}/v|{\vec{k}_2-\vec{q}}|)^2}}.\nonumber
\end{eqnarray}
The energy conservation in the delta function of Eq.~\eqref{eq:ee-rate} controls the allowed transitions. We split the delta function into two,
\begin{equation}
\delta(\eps_{\vec{k}_1}+\eps_{\vec{k}_2}-\eps_\vec{k_1+q}-\eps_\vec{k_2-q}) = \int d\omega \delta (\eps_{\vec{k}_1}-\eps_\vec{k_1+q}-\omega) \delta (\eps_{\vec{k}_2}-\eps_\vec{k_2-q}+\omega)
\end{equation}
With the aid of two delta functions, we change the product of Fermi distribution into a difference,
\begin{eqnarray}
f^\zero_{\vec{k}_1}(1-f^\zero_{\vec{k_1+q}})&=&\frac{f^\zero_{\vec{k}_1}-f^\zero_{\vec{k_1+q}}}{1-e^{\beta\omega}}\\
f^\zero_{\vec{k}_2}(1-f^\zero_{\vec{k_2-q}})&=&\frac{f^\zero_{\vec{k}_2}-f^\zero_{\vec{k_2-q}}}{1-e^{-\beta\omega}}
\end{eqnarray}
Now the e-e scattering rate becomes
\begin{eqnarray}
    \gamma_2
&=&
    4\pi  \beta \int \frac{d^2{\vec{k}_1}}{(2\pi)^2} \int \frac{d^2\vec k_2}{(2\pi)^2}
    \int \frac{d^2\vec q}{(2\pi)^2}
    \int d\omega \nn \\
&\times & \delta(\eps_{\vec{k}_1}-\eps_\vec{k_1+q}-\omega) \frac{v^2 k_1 (k_1+q \cos\phi)+\eps_{\vec{k}_1}\eps_\vec{k_1+q}}{v^2k_1|{\vec{k}_1+\vec{q}}|\sqrt{1+(\eps_{\vec{k}_1}/vk_1)^2} \sqrt{1+(\eps_{\vec{k}_1+\vec{q}}/v|{\vec{k}_1+\vec{q}}|)^2}}\nn
&\times & \delta (\eps_{\vec{k}_2}-\eps_\vec{k_2-q}+\omega)  \frac{v^2 k_2 (k_2-q \cos\phi_2)+\eps_{\vec{k}_2}\eps_\vec{k_2-q}}{v^2k_2|{\vec{k}_2-\vec{q}}|\sqrt{1+(\eps_{\vec{k}_2}/vk_2)^2} \sqrt{1+(\eps_{\vec{k}_2-\vec{q}}/v|{\vec{k}_2-\vec{q}}|)^2}} |V_q|^2  \nn
&\times& \frac{f^\zero_{\vec{k}_1}-f^\zero_{\vec{k_1+q}}}{1-e^{\beta\omega}} \frac{f^\zero_{\vec{k}_2}-f^\zero_{\vec{k_2-q}}}{1-e^{-\beta\omega}} 32\lb \lp \frac{q}{2k_F}\rp^2 - \lp \frac{q}{2k_F}\rp^4\rb, \label{eq:ee-rate2}
\end{eqnarray}
We shift the vector $\vec {p+q}\to -\vec p $ so that
\begin{eqnarray}
\int d^2{\vec{k}_1} \lp f^\zero_{{\vec{k}_1}} - f^\zero_{{\vec{k}_1+\vec{q}}}\rp \delta(\eps_{\vec{k}_1} -\eps_\vec{k_1+q}-\omega)\lp \dots\rp &=& \int d^2{\vec{k}_1}  \left[ f^\zero_{{\vec{k}_1}}\delta(\eps_{\vec{k}_1} -\eps_\vec{k_1+q}-\omega) - f^\zero_{-{\vec{k}_1} } \delta (\eps_\vec {-k_1-q} -\eps_\vec{-k_1}-\omega)\right] \lp \dots\rp \nn
&=& \int d^2{\vec{k}_1}  f^\zero_{{\vec{k}_1}}\lb \delta(\eps_{\vec{k}_1} -\eps_\vec{k_1+q}-\omega) -  \delta (\eps_{\vec{k}_1+\vec{q}} -\eps_{\vec{k}_1}-\omega)\rb \lp \dots\rp.
\end{eqnarray}
and to get the last line, we used the symmetry of $\eps_{\vec{k}_1}$ and $f^\zero_{{\vec{k}_1}}$. Similarly for ${\vec{k}_2}$, we get,
\begin{align}
\int d^2{\vec{k}_2} \lp f^\zero_{\vec{k}_2} - f^\zero_{{\vec{k}_2-\vec{q}}}\rp \delta(\eps_{\vec{k}_2} -\eps_\vec{k_2-q}+\omega)
\lp \dots\rp &=& \int d^2{\vec{k}_2}  f^\zero_{{\vec{k}_2}}\lb \delta(\eps_{\vec{k}_2} -\eps_\vec{k_2-q}+\omega) -  \delta (\eps_{\vec{k}_2-\vec{q}} -\eps_{\vec{k}_2}+\omega)\rb \lp \dots\rp.
\end{align}
Now we examine the energy conservation during the scattering process,
\begin{eqnarray}
\eps_{\vec{k}_1+\vec{q}} -\eps_{\vec{k}_1}&=& \sqrt{\lambda_R^2 + v^2 (k_1^2+q^2+2k_1q \cos \phi)} - \sqrt{\lambda_R^2 + (vk_1)^2}\nn
\eps_{\vec{k}_2-\vec{q}} -\eps_{\vec{k}_2}
&=& \sqrt{\lambda_R^2 + v^2 (k_2^2+q^2-2k_2q \cos \phi_2)} - \sqrt{\lambda_R^2 + (vk_2)^2}\label{2}
\end{eqnarray}
and perform the delta function integrations over angles,
\begin{eqnarray}
&&\int d\phi \ft G(\phi) \delta \lp \sqrt{\lambda_R^2 + v^2 (k_1^2+q^2+2k_1q \cos \phi)} - \sqrt{\lambda_R^2 + (vk_1)^2}+\omega\rp - (\omega \to -\omega) \nn
&&    =\sum_{j=1,2}\ft G(\phi_j^+) \left |\frac{|\sqrt{\lambda_R^2+ (vk_1)^2}-\omega|}{v^2 k_1 q \sin \phi_j^+ }\right| -\ft G(\phi_j^-)\left |\frac{|\sqrt{\lambda_R^2+ (vk_1)^2}+\omega|}{v^2 k_1 q \sin \phi_j^- }\right|,
\end{eqnarray}
where $\phi_j^\pm$ are the solutions that make the arguments in the delta function become zero
\begin{align}
\phi_j^+= &\pm \cos^{-1}\lp \frac{\omega^2-v^2 q^2 -2 \omega \sqrt{\lambda_R^2+(vk_1)^2}}{2v^2k_1q}\rp,\nn
\phi_j^-= &\pm \cos^{-1}\lp \frac{\omega^2-v^2 q^2 +2\omega \sqrt{\lambda_R^2+(vk_1)^2}}{2v^2k_1q}\rp.\label{eq:phi}
\end{align}

%%%%%%%%%%%%%%%%%%%%%%%%%%%%%%%%%%%%%%%%%%%%%%%%%%%%%%%%%%%%%%%
The form of Eq.~\eqref{eq:phi} makes the integration over $\omega$, $q$ and $p$ rather intricate. To simplify the expression further, firstly we assume that $\omega\propto T$ is very small compared to the Fermi energy.
Then, Eq.~\eqref{eq:phi} will become
\begin{eqnarray}
\cos \phi_j^\pm= - \frac{ q}{2k_1}
\end{eqnarray}
This simplifies the expression in Eq.~(\ref{eq:ee-rate2})
\begin{eqnarray}
{\vec{k}_1} &\approx& {\vec{k}_1+\vec{q}}
\lp -\lambda_R +\sqrt{\lambda_R^2 + v^2 (k_1^2+q^2+2k_1q \cos \phi)} \rp \nn
&=& \lp -\lambda_R +\sqrt{\lambda_R^2 + (vk_1)^2} \rp ^{2}
\end{eqnarray}
Performing the radial integration over $k_1$ yields,
\begin{eqnarray}
&&\int d k_1 k_1 f^\zero_{{\vec{k}_1}} \Biggl \{ \sum_{j=1,2} \ft G^\pm(\phi_j^+) \left |\frac{|\sqrt{\lambda_R^2+ (vk_1)^2}-\omega|}{v^2 k_1 q \sin \phi_j^+ }\right| -\ft G^\pm(\phi_j^-)\left |\frac{|\sqrt{\lambda_R^2+ (vk_1)^2}+\omega|}{v^2 k_1 q \sin \phi_j^- }\right| \Biggr \} \nn
&=&
\int_{q/2}^{k_F} d k_1 k_1 f^\zero_{{\vec{k}_1}} \ft G^\pm(\phi)  \frac{-4\omega}{v^2 k_1 q \sqrt{1-\displaystyle \lp \frac{q}{2k_1}\rp^2}  },\nn
&=& \frac{-4\omega k_F}{2q v^2}\left\{\sqrt{4-\tilde q^2}-\tilde q \displaystyle \lb \tan^{-1} \lp \frac{\sqrt{4-\tilde q^2}}{\tilde q}\rp\mp \tan^{-1} \lp \frac{\tilde \lambda_R}{\tilde q}\sqrt{\frac{4-\tilde q^2}{1+ \tilde \lambda_R^2}}\rp \rb \right\}\label{eq:radial}
\end{eqnarray}
where
\begin{eqnarray}
G^\pm(\phi) &=&\frac{v^2 k_1 (k_1+q \cos\phi)+\eps^\pm_{\vec{k}_1}\eps^\pm_\vec{k_1+q}}{v^2k_1|{\vec{k}_1+\vec{q}}|\sqrt{1+(\eps^\pm_{\vec{k}_1}/vk_1)^2} \sqrt{1+(\eps^\pm_{\vec{k}_1+\vec{q}}/v|{\vec{k}_1+\vec{q}}|)^2}}\nn
&=& 1- \frac{v^2 q^2 }{2\lb v^2k_1^2+\lp \mp\lambda_R +\sqrt{\lambda_R^2 + (vk_1)^2} \rp ^{2}\rb  },
\end{eqnarray}
and $\tilde q=q/k_F$, and  $\tilde \lambda_R=\lambda_R/(vk_F)$. The lower bound of the integral $k_1=q/2$ is placed so that $\sin\phi$ is well defined. The upper bound of the integral $k_1=k_F$ is due to the low-temperature limit $f^\zero_{\vec {p}}=\theta(\mu - \epsilon _{\vec{k}_1})$. The radial integration over $k_2$ gives the opposite sign of Eq.~\eqref{eq:radial}. The opposite signs of the $k_1$ and $k_2$ integrals will result in a positive value of $\omega$ integral,
\begin{eqnarray}
\int_{-\infty}^{\infty}  d\omega  \frac{-\omega^2}{(1-e^{\beta \omega})(1-e^{-\beta \omega})}
&=&\frac{2 \pi ^2}{3 \beta^3}
\end{eqnarray}
After performing the $\vec k_1$, $\vec k_2$ and $\omega$ integrals, we are left with a $q$ integral as follows,
\begin{eqnarray}
\gamma_{2}^\pm&=&\frac{4\pi}{\hbar}\frac{2\pi^2}{3\beta^2}\frac{4(2\pi)^3 e^4k_F^2}{\hbar^4 v^4(2\pi)^6} \int_0 ^{2k_F} q dq \frac{1}{q^2} \left\{\sqrt{4-\tilde q^2}-\tilde q \displaystyle \lb \tan^{-1} \lp \frac{\sqrt{4-\tilde q^2}}{\tilde q}\rp\mp \tan^{-1} \lp \frac{\tilde \lambda_R}{\tilde q}\sqrt{\frac{4-\tilde q^2}{1+ \tilde \lambda_R^2}}\rp \rb \right\}^2\nn
&& \times 32 \lb \lp \frac{q}{2k_F}\rp^2-\lp \frac{q}{2k_F}\rp^4\rb \frac{1}{(q+d)^2}.\nn
&=&\frac{32}{3} \frac{e^4 (k_B T)^2}{\hbar^5 v^4} \tilde \gamma^\pm(\tilde\lambda_R, \tilde d)\label{eq:gamma2}
\end{eqnarray}
We plot the viscosity as a function of $\lambda_R$ in Fig.~\ref{fig:gamma}(b).
\end{widetext}

\bibliography{refs}

\end{document}